\documentclass[a4paper,11pt]{article}
\pdfoutput=1 
\usepackage{jcappub}
\usepackage[utf8]{inputenc}
\usepackage{hyperref}
\usepackage{xcolor}
\usepackage{graphicx}
\usepackage{amsmath,amssymb,amsbsy,amstext,amsthm}
\usepackage{mathtools}
\usepackage{amsfonts}

\newenvironment{equations}{\equation\aligned}{\endaligned\endequation}

\def\beq{\begin{equation}}
\def\eeq{\end{equation}}
\def\ba{\begin{equations}}
	\def\ea{\end{equations}}
\def\bc{\begin{center}}
	\def\ec{\end{center}}

\def\n{\nabla}

\def\cl{{\cal L}}

\begin{document}
	\title{
        Doppelg{\"a}nger dark energy: modified gravity with non-universal couplings after GW170817\medskip}
	
    	\author[a]{{Luca Amendola,}}
		\emailAdd{{l.amendola}@{thphys.uni-heidelberg.de}}
    
    	\author[a]{{Dario Bettoni,}}
		\emailAdd{{bettoni}@{thphys.uni-heidelberg.de}}
    
		\author[a]{{Guillem Dom\`enech}}
		\emailAdd{{domenech}@{thphys.uni-heidelberg.de}}
	
		\author[b]{{and Adalto R. Gomes}}
		\emailAdd{{argomes.ufma}@{gmail.com}}
	
	%	\author{\textsc{$^{c}$} }
	%	\email{{}@{}}
	
	\affiliation[a]{{Institut f\"ur Theoretische Physik, Ruprecht-Karls-Universit\"at Heidelberg Philosophenweg 16, 69120 Heidelberg, Germany}}
	\affiliation[b]{{Departamento de F{\'i}sica, Universidade Federal do Maranh{\~a}o (UFMA)
Campus Universit{\'a}rio do Bacanga, 65085-580, S{\~a}o Lu{\'i}s, Maranh{\~a}o, Brazil}}

	\abstract{
    Gravitational Wave (GW) astronomy severely narrowed down the theoretical space for scalar-tensor theories. We propose a 
    {new} class of attractor models {for Horndeski action} in which GWs propagate at the speed of light in the nearby universe but not in the past. To do so we derive new solutions to the interacting dark sector in which the ratio of dark energy and dark matter remains constant, which we refer to as {\it doppelg{\"a}nger dark energy} (DDE). We then remove the interaction between dark matter and dark energy by a suitable change of variables.
The accelerated expansion that (we) baryons observe is due to a conformal coupling to the dark energy scalar field.
We show how in this context it is possible to find a non trivial subset of  solutions in which GWs propagate at the speed of light only at low red-shifts.
The model is an attractor, thus reaching the limit $c_{T}\to1$ relatively fast. However, the effect of baryons turns out to be non-negligible and severely constrains the form of the Lagrangian. In passing, we found that in the simplest DDE models the no-ghost conditions for perturbations require a non-universal coupling to gravity. In the end, we comment on possible ways to solve the lack of matter domination stage for DDE models.}

\keywords{Dark energy, scalar-tensor theories, gravitational waves}

\arxivnumber{1803.06368}
	
	\maketitle
	\section{Introduction}
	
    Since the discovery of the accelerated expansion of the Universe a vast class of dark energy models have been proposed. In other words, we still lack of a solid explanation for the mechanism behind it. Most dark energy models are basically identical to the standard cosmological model, a.k.a. $\Lambda$CDM, at the background evolution but might differ at the linear and non-linear perturbation level. Among these, scalar-tensor theories of gravity have played a pivotal role and have witnessed in recent years significant theoretical advances. The re-discovery of the most general scalar-tensor theory that gives second order equations of motion, Horndeski action \cite{Horndeski:1974wa} or Covariant Galileons \cite{Deffayet:2009wt}, and their extensions \cite{Zumalacarregui:2013pma,Gleyzes:2014qga,Gleyzes:2014dya,Crisostomi:2016czh,Crisostomi:2016tcp,Achour:2016rkg,Langlois:2015cwa} provided a very general framework for such theories. The drawback is that the theory space is extremely large and hard to constrain.
    
   The large degeneracy between dark energy models start to face with the reality of observations. In fact, most of these models predict an anomalous propagation speed for gravitational waves \cite{Lombriser:2015sxa,Lombriser:2016yzn,Bettoni:2016mij}. The almost simultaneous detection of GWs and the electromagnetic counterparts tells us that within $40$ Mpc (at $z\sim0.08$) from us GWs propagate at the speed of light \cite{LIGO:2017qsa}. Since the signals arrived within $1$s difference and light took $10^{15}$ s to reach us, we have that $|c^2_T/c^2-1|<10^{-15}$. Such tight constraint immediately ruled out most of the Dark Energy (DE) theories containing derivative couplings to gravity or at least those models which show this effect in the nearby universe (in cosmological scales) \cite{Ezquiaga:2017ekz,Creminelli:2017sry,Baker:2017hug,Langlois:2017dyl,Sakstein:2017xjx}. Nevertheless, the window for other dark energy models, e.g. with non-minimal couplings to gravity, non-local gravity, etc., is still large \cite{Crisostomi:2017pjs}. The situation becomes increasingly interesting if one considers interaction among dark energy and dark matter \cite{Kunz:2007rk,Bean:2008ac,Battye:2018ssx}. For example, see Ref.~\cite{DiValentino:2017iww} where interacting dark energy could provide a solution to the $H_0$ tension between Planck and local measurements and Ref.~\cite{Barros:2018efl} where it is used to solve the $\sigma_8$ tension. As we will see they also provide a way to avoid the GWs constraint within the Horndeski theory without considering any fine-tuning of the matter couplings or cancellations among Horndeski functions.
	
	On top of all that, the fact that the energy density of Dark Matter (DM) and DE are so close at present eludes explanation. The so-called coincidence problem could be alleviated if the energy density of DE is proportional to the energy density of DM and this proportionality is constant in time in the nearby universe. The coincidence problem is then set aside to a order-of-unity constant which must be fixed by observations. This mechanism was first proposed in Ref.~\cite{Amendola:1999qq} using interacting DE and we will refer to these solutions as Doppelg{\"a}nger\footnote{Doppelg{\"a}nger, from the German word for lookalike, refers here to the property that DE behaves like matter but is not identical with it.} Dark Energy (DDE), thus avoiding the use of terms like scaling or tracking solutions, that have been applied also to different set ups, e.g. Ref.~\cite{Crisostomi:2017pjs}. More DDE solutions have been found in Refs.~\cite{Amendola:2006qi,Gomes:2013ema,Gomes:2015dhl} in the context of scalar-tensor (Horndeski) theories. Interestingly, these solutions do not only have applications to the late universe but have been used in different situations, e.g. in primordial black hole scenarios \cite{Amendola:2017xhl}, growing matter cosmology \cite{Amendola:2007yx}, etc.
    
  In DDE models, DE and DM interact in such a way so that both components behave as a single fluid with an effective equation of state. This implies that the DE Lagrangian has to be of a specific form, compatible with the modified evolution of DM. At the perturbation level, they will obviously behave differently. There are, however, two drawbacks in this approach. First, the functional form of the DE Lagrangian depends on the form of the interaction with DM. Thus, finding a general solution is a non-trivial task \cite{Gomes:2015dhl}. This methodology works well for K-essence models but gets increasingly complicated with Horndeski Lagrangians (and beyond)\cite{Gomes:2015dhl}. Secondly, the DDE accelerating solution is present as soon as DM dominates and, since it is an attractor solution, the system relaxes there relatively fast. For this same reason, DDE usually lacks of an epoch of regular matter domination \cite{Amendola:2006qi}. The usual way out is to consider a baryon dominated stage or that DE is doppelg{\"a}nger of neutrinos instead of DM \cite{Amendola:2007yx}.  We will suggest alternative solutions to the shortcomings discussed above.
 
 In this work, we propose a new way to approach DDE in general scalar-tensor theories and find a more general DDE action, extending previous results. Furthermore, we investigate the effects of the recent constraint on the speed of gravitational waves on the DDE action. We found that DDE solutions can be made compatible with the recent constraints while still allowing for non-trivial effects out of the DDE regime. In our approach, we first remove the DM and DE interaction by a conformal transformation of the metric. Once we work in the newly defined metric -- usually referred to as working in a different frame -- the requirements for DDE are straightforward and the lengthy process to find solutions is simplified. In this new picture, the energy density of DE just happens to behave like a matter fluid. The acceleration of the universe observed by baryons is then due to a conformal coupling between baryons and DE. Neglecting the effects of baryons for the background evolution, we find the most general solutions of DDE. 
 
 Lastly, treating baryons as a perturbation to our solutions, we find that baryons tend to take the system out of the DDE attractor by $1\%$ at linear level. While this has no important impact on the background evolution nor on scalar perturbations, one expects that a $1\%$ deviation from $c_T^2=1$ is ruled out by observations. We use the GWs observation to constrain the form of the Lagrangian. This also has significant implications for fine-tuned models, in which the fine-tuning is chosen in the absence of matter fields. We thus expect that either the fine-tuned model would be ruled out when one takes into account matter fields or it should be further fine-tuned to account for such deviation \cite{Baker:2017hug,Langlois:2017dyl}. The advantage of using DDE models is that we only have to consider the deviations due to baryons, as DM and DE behave as a single fluid.

	The paper is organized as follows. In Sec.~\ref{sec:interacting}, we review the previous approach to DDE solutions and we show that there always exists a frame in which the interaction between DM and DE is absent. We place emphasis on what are the observables and how they do not depend on the frame. In Sec.~\ref{sec:solutions} we proceed to find the DE Lagrangian compatible with DDE solutions. We do so using a different approach than in Ref.~\cite{Amendola:1999qq,Gomes:2013ema,Gomes:2015dhl}, namely we focus on the rough behaviour of the energy density of DE, similar in spirit to Ref.~\cite{DeFelice:2011bh}. We find new solutions and provide a way to study the phase space in complete generality. In Sec.~\ref{sec:aplications}, we study a particular case to model the current acceleration of the universe and compatible with the recent GW observation. At the end of this section, we provide a way to have a matter dominated stage in DDE. We further discuss about possible screenings and the behavior of this solution during matter and radiation domination. We conclude our work in Sec.~\ref{sec:conclusions}. Explicit formulas can be found in the Appendices.

	\section{Interacting dark matter and metric transformations\label{sec:interacting}}
	
A key ingredient to get naturally accelerating DDE solutions seems to be a non-trivial interaction between DM and the scalar field responsible for DE. This is readily seen from the fact that if there is no interaction and the fields are minimally coupled to gravity, the only option to get a proportionality between matter and DE energy densities is that the scalar field behaves like dust. Clearly, this cannot describe the current expansion of the universe. If one uses a non-trivial interaction between DM and DE, leaving the Standard Model (SM) sector uncoupled, then the effective equation of state of dark matter is modified and both DM and DE behave on the background as a single fluid with a single effective equation of state on cosmological scales. Note that the small scales behaviour will be clearly different. Usually such interaction is modelled at the level of the equations of motion by a term violating the energy conservation of DM. In a Friedmann–Lema\^itre–Robertson–Walker (FLRW) metric, the interaction reads
		\ba\label{eq:dmcons}
		\frac{d\bar \rho_{DM}}{d\bar t}+3\bar H\bar \rho_{DM}=Q(\phi)\frac{d\phi}{d\bar t}\bar \rho_{DM}\,,
		\ea
	where $\bar t$ is the cosmic time, $\bar \rho_{DM}$ is the energy density of dark matter, $\bar H$ is the expansion parameter, $\phi$ is the DE scalar field and  $Q(\phi)$ is an arbitrary function of the scalar field.
    The main difficulty to find a general DDE Lagrangian 
    is that the function $Q(\phi)$ needs to appear inside the DE Lagrangian functions and significantly complicates the analysis. If the DE Lagrangian contains non-minimal and derivative couplings to gravity, then it becomes extremely involved -- even if one assumes that $Q$ is a constant \cite{Gomes:2015dhl}. By removing the interaction, we will avoid some of the complications.

	To be more clear on this statement, let us work in the action formalism. The action can be written in the following form,
		\ba\label{eq:baraction}
		S=\int d^4x \sqrt{-\bar g} \left\{\sum_i \bar{\cl}_i(\bar g,\phi)+\bar\cl_{DM}\left(\phi\right)+\bar\cl_{SM}\right\}\,,
		\ea
	where the DE Lagrangian $\bar{\cl}_i$'s are given by the Horndeski Lagrangian \cite{Deffayet:2009wt} (shown explicit later), $\bar {\cl}_{DM}$ is the Lagrangian for dark matter and $\bar {\cl}_{SM}$ is the standard model Lagrangian (for our purposes baryons and radiation). Note that baryons and radiation are minimally coupled to the metric $\bar g_{\mu\nu}$ and, thus, we call this form of the action the \textit{matter frame}, which need not coincide with the Einstein frame -- gravity is not necessarily given by GR. We model the DM Lagrangian by
		\ba\label{eq:inter}
		\bar \cl_{DM}=-\frac{\lambda}{2}\left(\bar \n_\mu\sigma\bar \n^\mu\sigma + B^{-2}(\phi)\right)\,,
		\ea
 where $\lambda$ is a Lagrange multiplier,\footnote{Any dependence on $\phi$ in front of the Lagrange multiplier $\lambda$ does not have any practical effect.} $B(\phi)$ is a non-zero well behaved function of $\phi$ and $\bar\nabla_\mu\sigma$ is the 4-velocity of the DM fluid. This form of the Lagrangian\footnote{Note that a potential for $\sigma$ would also give dust.} is known to give a dust fluid for $B=1$ \cite{Lim:2010yk,Bettoni:2012xv} and it is widely used in mimetic gravity \cite{Chamseddine:2013kea}. Such kind of non-minimal couplings between a scalar field and matter field is ubiquitous in higher dimensional theories, e.g. string theory and braneworld, and in $R^2$ models \cite{Fujii:2003pa,Piazza:2004df}. It often takes the form of ${\rm e}^{q\varphi}$ where $q$ is related to the parameters of the underlying theory and is referred to as dilatonic coupling. It should be noted that in the present case DM and SM are non-universally coupled to gravity. It would be interesting to derive this kind of non-universal coupling from a fundamental set up. This could probably be realized in a braneworlds, where the interaction of DM with the extra dimension is different to that of baryons \cite{KOIVISTO:2013jwa}; similar to the inflationary model in Ref.~\cite{Larrouturou:2016lzr}, where the metric is different if scalar field lives in the bulk or in the brane.
   
To illustrate the interaction, let us focus on a FLRW background given by
\ba
d\bar s^2=-\bar N ^2d\bar t^2+\bar a^2(\bar t)\delta_{ij}dx^idx^j\,,
\ea
where $\bar N$ is the Lapse function and $\bar a$ is the scale factor. Variation of the action \eqref{eq:baraction} with respect to the Lagrange multiplier $\lambda$ yields ${d\sigma}/{d\bar t}=B^{-1}$. One can then see that the energy density of the dust fluid is given by $\bar\rho_{DM}=\lambda/B^2$. In this way, varying the action with respect to $\sigma$, one recovers Eq.~\eqref{eq:dmcons} with $\bar H={d\ln \bar a}/{d\bar t}$  and the identification
\ba
Q\equiv -\frac{d\ln B}{d\phi}\,.
\ea
The Friedman equations are given by
\ba\label{eq:fried1}
\sum_{i=2}^5\bar{\cal E}_i+\bar\rho_{DM}+\bar\rho_{b}+\bar\rho_{rad}=0\,
\quad{\rm and}\quad
\sum_{i=2}^5\bar{\cal P}_i+\bar p_{rad}=0\,.
\ea
where we included baryons and radiation and we defined $\bar{\cal E}_i\equiv{-}{\bar a^{-3}}\tfrac{\delta }{\delta \bar N}\bar{\cl}_{i}|_{ \bar N=1}$ and $\bar{\cal P}_i\equiv\tfrac{\bar a^{-2}}{3}\tfrac{\delta}{\delta \bar a}\bar{\cl}_{i}|_{\bar N=1}$ as in Ref.~\cite{DeFelice:2011hq}. 
A quick inspection to Eq.~\eqref{eq:inter} tells us that the interaction between DE and DM, i.e. the function $B$, can be absorbed into the metric $\bar g_{\mu\nu}$. Therefore, we can work in a frame -- in a new metric -- where DE and DM do not interact. It is important to note that this is always possible and independent of the functional form of $B$.

	\subsection{Removing interactions by a change of variables}
	
In order to achieve the desired frame change, we inspect Eq.~\eqref{eq:inter} and notice that the DM 4-velocity is geodesic of the metric
		\ba\label{eq:DMtoSM}
		g_{\mu\nu}=B^{-2} \bar g_{\mu\nu}\,.
		\ea
	We can thus rewrite our action in terms of the new conformal metric in which DM behaves as the usual pressure-less fluid. The DM Lagrangian in the new frame is explicitly given by
	\ba\label{eq:DM}
	\cl_{DM}=-\frac{\lambda}{2} B^{2}\left(\n_\mu\sigma\n^\mu\sigma + 1\right)\,.
	\ea
	In this form it is clear that we have a pressureless fluid with conserved energy density. The new FLRW metric reads
	\ba
	ds^2=-N^2dt^2+a^2(t)\delta_{ij}dx^i dx^j\,,
	\ea
	where
		\ba
		a = B^{-1}\bar a \quad {\rm and}\quad  dt=B^{-1} d\bar t\,.
		\ea
		Note that $\bar N=N$ since we have already redefined the time coordinate at the background level. We can use the same logic as before to find that the energy density of DM in this frame is $\rho_{DM}\equiv{\lambda}  B^{2}$ and it satisfies
	\ba\label{eq:rhodm}
	\dot\rho_{DM}+3 H \rho_{DM}=0\,,
	\ea
	where $\dot ~\equiv d/dt$ and $H\equiv \dot a /a $. If we did a similar exercise but for a general fluid $I$ with interaction $Q_I$ with DE in the matter frame, we would find that $\bar\rho_I = B^4 \rho_I$, $\bar p_I = B^4 p_I$, $\bar w_I = w_I$ and 
	\ba
	\dot\rho_I + 3 H \left(1+w_I\right)=\left(\frac{d\ln B}{d\phi}\left(3w_I-1\right)+Q_I\right)\dot\phi \,\rho_I\,.
	\ea
	Recall that radiation ($w_r=1/3$) is conformal invariant. In this new frame baryons will now get a coupling to dark energy but since we are interested in recent epochs where baryons are subdominant we neglect them for now. However, as we shall explore later, this component plays nonetheless an important role. 
Since in this frame DM is minimally coupled to the metric $g_{\mu\nu}$ we call the corresponding form of the action the \textit{DM frame}. Let us emphasize that ``barred'' quantities always refer to the matter frame and ``unbarred'' ones to the dark matter frame. On the other hand, the DE Lagrangian transforms as well and the action is given by
	\ba
	S=\int d^4x \sqrt{- g} \left\{\sum_i {\cl}_i( g,\phi)+\cl_{DM}+\cl_{SM}(\phi)\right\}\,.
	\ea
	The relation between $\bar{\cl}_i$'s and ${\cl}_i$'s up to ${\cal L}_4$ can be found in the App.~\ref{app:rules} (see also Refs.~\cite{Bettoni:2013diz,Crisostomi:2016tcp,Achour:2016rkg}). The important point is that the dependence on $B$ appears on ${\cl }_{SM}$ and in ${\cl}_i$'s. Nevertheless, since we consider the effect of baryons and radiation to be irrelevant as a first order approximation, the particular form of $B$ in ${\cl}_i$'s is irrelevant in the DM frame at first order approximation as we will treat the ${\cl}_i$'s as general as possible. 
	
    Before going into the details of the solutions, it is important to review what are the physical observables. It is well-known that physics should not depend on field redefinitions; for the case of gravity see for example Ref.~\cite{Deruelle:2010ht}. In late time cosmology one uses the redshift and the luminosity distance relation. In the presence of a general non-minimal coupling of the DE scalar to baryons -- certainly the case of the DM frame -- we find that the luminosity distance relation is given by \cite{Deruelle:2010ht} (also see App.~\ref{app:rules})
	\ba
	D_L=\left(1+z\right)\int dz\frac{B}{H(z)\left(1+ \frac{d\ln B}{dN}\right)}\,,
	\ea
	where $dN=Hdt$ and $z$ is the redshift. Thus, observations only tells us about the combined effect of the matter energy momentum tensor and the non-minimal coupling. In order to extract more information we need to make further assumptions. For example, for $\Lambda$CDM we assume that there is no interaction and that DM is a pressure-less fluid. For interacting dark sector model, we face a dark degeneracy \cite{Kunz:2007rk}, i.e., we cannot distinguish the effects of DM and DE and, therefore, we cannot tell DM and DE apart.
    
    Note that most of the calculations in the literature are done in the matter frame, i.e. where the SM is uncoupled. Therefore, for an easier comparison, we shall show the relation between quantities in both frames. First, the Hubble parameters are related by
	\ba\label{eq:hubble}
	H =B\bar H (1-\beta)\quad{\rm where}\quad \beta\equiv\frac{d\ln B}{d\bar N}\,,
	\ea
	and $d\bar N=\bar H d\bar t$. The effective equations of state are defined by
    \ba\label{eq:weff1}
	1+\bar w_{\rm eff}\equiv-\frac{2}{3 \bar H^2 }\frac{d\bar H}{d \bar t}\quad,\quad 1+ w_{\rm eff}\equiv-\frac{2}{3 H^2 }\frac{dH}{dt}
    \ea
    and are related by
	\ba\label{eq:weff2}
	1+w_{\rm eff}=\frac{1+\bar w_{\rm eff}-\frac23\beta-\frac{2}{3}\frac{d\ln (1-\beta)}{d\bar N}}{1-\beta}\,.
	\ea
	Note a couple of interesting things. First, there is no a priori bound on $w_{\rm eff}$ as $\beta$ is a free parameter. Second, only if ${d\ln \beta}/{d\bar N}= 0$ a constant effective equation of state will remain constant in any frame. The DE-DM proportionality constant will also depend on the frame and, hence, on $\beta$. Only if ${d\ln \beta}/{d\bar N}= 0$ the ratio will be constant in both frames, as we shall see in the next section. For these reason, we shall consider this case in what follows.
	
Let us end this section by giving an interpretation of the value of $\beta$ by assuming that $\bar w_{\rm eff}$ and $\beta$ are constant. If one assumes a power-law universe, certainly the case for a single barotropic fluid with constant equation of state,
 we have that $\bar H^2\propto \bar a^{-3\left(1+\bar w_{\rm eff}\right)}$ and $B\propto \bar a^ \beta$. We see that the effect of the conformal transformation is to change the expansion rate of the universe. For example, looking at \eqref{eq:hubble} we see that if $\beta>1$ then $H<0$ if $\bar H>0$ and vice-versa. So that we could go from a expanding universe to a contracting one \cite{Wetterich:2014eaa,Domenech:2015qoa}. The case $\beta=1$ (at all times) corresponds to Minkowski space. 

	\section{New Lagrangian with DDE solutions\label{sec:solutions}}
	
	The advantage of working in the DM frame is that we do not have to worry of the specific form of $Q$ and the DDE condition reduces only to find a DE Lagrangian that behaves as a pressurless fluid. Let us now focus on the DE Lagrangian. We will take the Horndeski form which is given by \cite{Deffayet:2009wt}
		\ba
		\cl_2&=G_2(\phi,X)\qquad{,}\qquad
		\cl_3= - G_3(\phi,X)\Box\phi\,,\\
		\cl_4&=G_4(\phi,X) R+G_{4,X}\left[\left(\Box\phi\right)^2-\n_\mu\n_\nu\phi\n^\mu\n^\nu\phi\right]\,,\\
		\cl_5&=G_5(\phi,X)G^{\mu\nu}\n_\mu\phi\n_\mu\phi-\frac{1}{6}G_{5,X}\left(\left(\Box\phi\right)^3-3\Box\phi\n_\sigma\n_\rho\phi\n^\sigma\n^\rho\phi+2\n_\sigma\n_\rho\phi\n^\sigma\n^\mu\phi\n_\mu\n^\rho\phi\right)\,,
		\ea
		where $X\equiv-\tfrac{1}{2}\n_\mu\phi\n^\mu\phi$ and $G_i$ with $i=2,3,4,5$ are general functions of $\phi$ and $X$. We did not include Beyond Horndeski terms for simplicity but the generalization is straightforward. As before the Friedman equations read
			\ba\label{eq:fried11}
			\sum_{i=2}^5{\cal E}_i+\rho_{DM}+\rho_{b}+\rho_{rad}=0
			\quad {\rm and}\quad
			\sum_{i=2}^5{\cal P}_i+p_{rad}=0\,,
			\ea
	where \cite{DeFelice:2011hq} ${\cal E}_i\equiv{-}{ a^{-3}}\tfrac{\delta }{\delta  N}{\cl}_{i}|_{  N=1}$ and ${\cal P}_i\equiv\tfrac{ a^{-2}}{3}\tfrac{\delta}{\delta  a}{\cl}_{i}|_{ N=1}$. The explicit forms can be found in App.~\ref{app:formulas}. For the moment we are only interested in the first Friedman equation given by \cite{DeFelice:2011bh}
	\ba\label{eq:1stfried}
	6H^2G_4=&\rho_\phi+\rho_{DM}+\rho_{b}+\rho_{rad}\,,
	\ea
	where
	\ba
	\rho_{\phi}\equiv&2XG_{2,X}-G_2+6X\dot\phi HG_{3,X}-2XG_{3,\phi}+24H^2X\left(G_{4,X}+XG_{4,XX}\right)-12HX\dot\phi G_{4,\phi X}\\&-6H\dot\phi G_{4,\phi}+2H^3X\dot\phi\left(5G_{5,X}+2XG_{5,XX}\right)-6H^2X\left(3G_{5,\phi}+2XG_{5,\phi X}\right)\,.
	\ea   
    
The DDE solutions are characterized by a constant ratio between $\rho_\phi$ and $\rho_{DM}$, namely we must require that
	\ba\label{eq:rhophi}
	\frac{d\ln\rho_{\phi}}{dN}=\frac{d\ln\rho_{DM}}{dN}=-3\,,
	\ea
	where in the last step we used Eq.~\eqref{eq:rhodm}. If we neglect baryons and radiation, i.e., $\rho_b=\rho_{\rm rad}=0$, the time derivative of Eq.~\eqref{eq:1stfried} yields
	\ba\label{eq:g4}
	\frac{d\ln G_4}{dN}=\frac{d\ln G_4}{d\ln \phi}\frac{d\ln \phi}{dN}+\frac{d\ln G_4}{d\ln X}\frac{d\ln X}{dN}=3w_{\rm eff}\,,
	\ea
    where we used Eq.~\eqref{eq:rhophi} and the definition $w_{\rm eff}$, Eq.~\eqref{eq:weff1}.
     It is not surprising that even if $\rho_{\phi}\propto \rho_{DM}\propto a^{-3}$ we have that $w_{\rm eff}\neq 0$ due to the presence of a non-minimal coupling.     To see this it is enough to use equation \eqref{eq:1stfried} and the $w_{\rm eff}$ definition. This gives $w_{\rm eff} \propto d(\ln G_4)/dN$. This equation already tells us how the system should behave. 
     
     Let us assume that $w_{\rm eff}$ is constant, which will be true if baryons and radiation are negligible or if we are in the adiabatic regime where ${d\ln w_{\rm eff}}/{dN}\ll 1$. To proceed further we have to solve for the dynamics of the scalar field. We can take another approach nonetheless. We will assume that $B$ is a dilatonic type coupling given by
    \ba\label{eq:B}
    B=\phi^q\,,
    \ea
   where $q$ is related to $\beta$ once the dynamics of $\phi$ are known. This functional form is the well-known dilatonic coupling in higher dimensional theories \cite{Fujii:2003pa} if one uses a field redefinition $\varphi\equiv\ln \phi$. Then the assumption that $\beta={\rm cnt}$ (see Eqs.~\eqref{eq:hubble} and \eqref{eq:weff2}) tells us that
	\ba\label{eq:lnp}
	\frac{d\ln B}{dN}=q\frac{d\ln \phi}{dN}= \frac{\beta}{1-\beta}={\rm cnt}
	\quad
	{\rm so\,\, that} 
	\quad
	\frac{d\ln \phi}{dN}\equiv\alpha={\rm cnt}\,,
	\ea
	where we used that $dN=(1-\beta) d\bar N$. It should be noted that the crucial assumption is that $\beta={\rm cnt}$ rather than the specific form of $B$ in Eq.~\eqref{eq:B}. In other words, if $\beta={\rm cnt}$ we can always find a field redefinition of $\phi$ where $B=\phi^q$. With these assumption, Eq.~\eqref{eq:lnp} also tells us that
	\ba\label{eq:lnx}
		\frac{d\ln X}{dN}=2\alpha-3\left(1+w_{\rm eff}\right)\,.
	\ea
	Using Eqs.~\eqref{eq:g4}, \eqref{eq:lnp} and \eqref{eq:lnx} we conclude that $G_4$ has to be a power law of $\phi$ and $X$. In fact, we can easily build any Horndeski function $G_i$ by noting that there is a constant combination, namely
	\ba
	Y\equiv X\phi^{p}={\rm cnt}\quad{\rm where}\quad p\equiv\frac{3}{\alpha}\left(1+w_{\rm eff}\right)-{2}\,,
	\ea
	which would not contribute to Eqs.~\eqref{eq:rhophi} and \eqref{eq:g4}. Thus, we can write in general that
	\ba
	G_i(\phi, X)=\phi^{p_i}a_i(Y) \qquad (i=2,3,4,5)\,.
	\ea
	We could have also used $X^{q_i}$ instead of $\phi^{p_i}$ but this is related by $q_i=p_i/2p$ and redefining a new function $ \tilde a_i(Y)\equiv Y^{p_i/2p} a_i(Y)$. We are left to find the relations among $p_i$'s and $p$ which are compatible with Eqs.~\eqref{eq:rhophi} and \eqref{eq:g4}. The latter straightforwardly gives
	\ba
	p_4=\frac{3}{\alpha}w_{eff}
	\ea
	A quick inspection to Eq.~\eqref{eq:rhophi} tells us that
		\ba
		G_2\propto a^{-3(1+w_{DM})}\Rightarrow p_2=-\frac{3}{\alpha}.
		\ea
	We can regard $p_2$, $p_4$ as the free parameters that determine $\alpha$ and $w_{\rm eff}$.     The remaining functions have to scale as
		\ba
		G_3\propto a^{3w_{\rm eff}-\alpha}\quad{\rm and}\quad G_5\propto a^{3+6w_{\rm eff}-\alpha}\,,
		\ea
	which imply
	\ba
	p_3=p_4-1\quad{\rm and}\quad p_5=2p_4-p_2-1\,.
	\ea
	This completes the general Lagrangian which admits DDE solutions. For a comparison with the literature we can derive a relation between equations of state given by
		\ba\label{eq:comparison}
		w_{\rm eff}=w_\phi\Omega_\phi\,,
		\ea
		 where we defined $\Omega_\phi=\rho_\phi/(6H^2G_4)$ and we used the time derivative of the first Friedman and the second Friedman equations, namely
		\ba
		-2\left(3H^2+2\dot H\right)G_4=p_\phi+p_{rad}\,,
		\ea
		where
		\ba
		p_\phi=&	G_2-2X\left(G_{3,\phi}+\ddot\phi G_{3,X}\right)
		-\left(4X\left(3H^2+2\dot H\right)+4H\dot X\right)G_{4,X}
		-8HX\dot X G_{4,XX}\\&+2\left(\ddot\phi+2H\dot\phi\right)G_{4,\phi}+4XG_{4,\phi\phi}+4X\left(\ddot\phi-2H\dot\phi\right)G_{4,\phi X }-4H^2X^2\ddot\phi G_{5,XX}\\&-2X\left(2H^3\dot\phi+2H\dot H\dot\phi+3H^2\ddot\phi\right)G_{5,X}+4HX\left(\dot X- HX\right)G_{5,\phi X}\\&+2\left(2\dot H X+2\dot X H +3H^2X\right)G_{5,\phi}+4HX\dot\phi G_{5,\phi\phi}\,.
		\ea
	and $w_\phi=p_\phi/\rho_\phi$ would be the equation of state for $\rho_\phi$. Note that $w_{\rm \phi}\neq 0$ as $\rho_{\phi}$ is not conserved. Interestingly, Eq.~\eqref{eq:comparison} is the same formula found in Ref.~\cite{Tsujikawa:2004dp}.
	
	Let us summarize the new solution to DDE Lagrangian. We have found that the Horndeski Lagrangian coefficient functions given by
		\ba\label{eq:trackerlagrangian}
		G_2(\phi,X)&=a_2(Y){\phi}^{p_2}\quad,\quad
		G_3(\phi,X)=a_3(Y){\phi}^{p_3}\,,\\
		G_4(\phi,X)&=a_4(Y){\phi}^{p_4}\quad,\quad
		G_5(\phi,X)=a_5(Y){\phi}^{p_5}\,,
		\ea
		where
		\ba
			Y=X\phi^{p}\quad, \quad p=p_4-p_2-2\quad,\quad p_3=p_4-1\quad{\rm and}\quad p_5=2p_4-p_2-1\,,
		\ea
		admit solutions where $\rho_\phi\propto\rho_{DM}$. 	Note that this form is a necessary condition to have DDE solutions. In order to be sufficient, there needs to be a relation among the free functions $a_i$. This will be found by imposing that, in absence of radiation, they satisfy $\sum_i {\cal P}_i=0$. For example, we can isolate $a_2$ in terms of the other functions. We would like to mention that the form of the Lagrangian reminds us of the tracker solutions found in \cite{DeFelice:2011bh} where it is required that $H\dot\phi^{2p}={\rm cnt}$. In our case it is $H^2\phi^{p_4}a^3={\rm cnt}$. Although different in practice, the spirit is similar.
        
        For later use we shall define here the DE-DM ratio in the dark matter and matter frame respectively as
    \ba
    c\equiv\frac{ \rho_{\phi}}{ \rho_{DM}}\quad{\rm and}\quad \bar c\equiv\frac{ \bar \rho_{\phi}}{ \bar\rho_{DM}}\,,
    \ea
    where $\bar\rho_\phi$ is defined as $\rho_{\phi}$ in Eq.~\eqref{eq:rhophi} but with the matter frame Horndeski functions $\bar G_i$. The DE-DM ratios are related by
    \ba\label{eq:ctobarc}
1+c=\left(1+\bar c\right){\left(1-\beta\right)^2}\,,
\ea
where we used Eqs.~\eqref{eq:hubble} and \eqref{eq:1stfried}. In this form one clearly sees that only if $\beta$ is constant both ratios can be constant at the same time. In what follows we shall assume that $\bar c$ and $\bar w_{\rm eff}$ take the same values as $\Lambda$CDM at the present time. Also, since we are, for the moment, treating $\beta$ as a free parameter it is convenient to impose first $\bar w_{\rm eff}\approx -0.7$ and $\bar c\approx 2.3$ and then use Eqs.~\eqref{eq:weff2} and \eqref{eq:ctobarc} to express $c$ and $w_{\rm eff}$ as functions of $\beta$.
		
	\subsection{Comparison with previous models\label{subsec:comparison}}
	For completeness we will compare our results with existing models in the literature. To do that we shall go back to the matter frame by undoing the conformal transformation Eq.~\eqref{eq:DMtoSM}. In this section we examine two illustrative cases. The explicit formulas are given in App.~\ref{app:rules}. It should be noted that we are assuming a dilatonic type coupling for $B$ and, therefore, the matter frame Lagrangian that we will obtain is only valid for such kind of interaction. However, it is important to emphasize that the solutions in the dark matter frame do not depend on the form of the coupling and the matter frame for a general coupling can be straightforwardly found. For an easy comparison with the literature we will keep our assumption that $B=\phi^q$ with $q$ a free parameter.
    
	In the first example, let us consider that $G_3=G_5=0$ and $G_4=\tfrac{1}{2}M^2_{pl}\phi^{p_4}$. In this case we find
		\ba
		\bar G_4=B^{-2}G_4\quad{\rm and}\quad
		\bar G_2= B^{-4}G_2+24X{B_{\phi}^2}\bar G_4\,.
		\ea
Additionally we require that $B=\phi^{p_4/2}$ so that $\bar G_4=\tfrac{1}{2}M^2_{pl}$. After a short algebra we get
		\ba
		\bar G_2
        \equiv\bar X\phi^{-2} g(Y)
		\ea
		where
		\ba
		g(Y)=\frac{a_2(Y)}{Y}+3p^2_4 M_{pl}^2\quad{\rm and}\quad Y=\bar X\phi^{2p_4-p_2-2}\,.
		\ea
		This form will look more familiar after a field redefinition $\varphi=\ln\phi$. In this notation
		\ba
		\bar G_2= \bar X_{\varphi} g(\bar X_{\varphi}{\rm e}^{\lambda\varphi}) \quad{\rm where}\quad \lambda =2p_4-p_2\,
		\ea
		and $\bar X_\varphi\equiv-\tfrac{1}{2}\bar\n_\mu\varphi\bar\n^\mu\varphi$. This recovers the very well known form of DDE solutions \cite{Tsujikawa:2004dp,Amendola:2006qi}.
		
		In our second example, let us briefly expand the previous case to include $G_3$. Using the same assumptions on $B$ and $G_4$ we find
		\ba
		 \bar G_3=B^{-2}G_3-2 B^{-2} G_{4}\frac{B_\phi}{B}
		\ea
		which yields
		\ba
		\bar G_3=\phi^{-1} \left(a_3(Y)-qM_{pl}^2\right)\,.
		\ea
		In the action this terms appears as $\bar G_3\bar\Box\phi$. Thus doing the field redefinition $\varphi=\ln\phi$ we find that
		\ba\label{eq:actiong3}
		S\subset -\int d^4x \sqrt{-\bar g} \,\bar a_3(Y)\bar\Box\varphi\quad{\rm where}\quad \bar a_3(Y)\equiv a_3(Y)-qM_{pl}^2\,.
		\ea
		Note that the last term in the right hand side is just a constant and thus yields a total derivative.
        
		In general, the Lagrangian in the matter frame, where most of the literature works with, is given by
		\ba\label{eq:matterframelagrangian}
		\bar G_2(\varphi,\bar X_{\varphi})&={\rm e}^{{\bar p_2} {\varphi}} \bar a_2(\bar X_{\varphi}{\rm e}^{\lambda \varphi})\quad,\quad
		\bar G_3(\varphi,\bar X_{\varphi})={\rm e}^{{\bar p_3} {\varphi}}\bar a_3(\bar X_{\varphi}{\rm e}^{\lambda \varphi})\\
		\bar G_4(\varphi,\bar X_{\varphi})&={\rm e}^{{\bar p_4} {\varphi}}\bar a_4(\bar X_{\varphi}{\rm e}^{\lambda \varphi})\quad,\quad
		\bar G_5(\varphi,\bar X_{\varphi})={\rm e}^{{\bar p_5} {\varphi}}\bar a_5( \bar X_{\varphi}{\rm e}^{\lambda \varphi})
		\ea
		where
		\ba
		\lambda=\bar p_4-\bar p_2\quad,\quad \bar p_3=\bar p_4\quad{\rm and}\quad \bar p_5=2\bar p_4-\bar p_2\,.
		\ea
        The relation with the dark matter frame exponents are 
		\ba
		\bar p_4=p_4-2q \quad{\rm and} \quad\bar p_2=p_2-4q.
		\ea
		Note that we are working with $\varphi$ and therefore the form of $\bar G_3$ and $\bar G_5$ differ by a factor $\phi={\rm e}^{\varphi}$ when using $\phi$ instead. This Lagrangian has to be supplied with the interaction with DM that is given by
		\ba
		\frac{d\bar \rho_{DM}}{d\bar t}+3\bar H\bar \rho_{DM}=-q\frac{d\varphi}{d\bar t}\bar \rho_{DM}\,.
		\ea
		The effective equation of state is given by
		\ba
		\bar w_{\rm eff}=-\frac{\bar p_4+q}{\bar p_2+q}\,.
		\ea
		Note that for $\bar p_4=\bar p_2$ we have $\bar w_{\rm eff}=-1$. For $\bar p_4=0$ we have
		$
		\bar w_{\rm eff}=\frac{q}{q-\lambda}\,,
		$
		which is exactly what Ref.~\cite{Tsujikawa:2004dp} finds.
		As one can see, working in the matter frame involves the quantity $q$ in the DM-DE system. The advantage of working in the dark matter frame is that $q$ is not present and thus we can draw general results more clearly.
        
        Our results go beyond that found in Refs.~\cite{Gomes:2013ema,Gomes:2015dhl}. The first reason for our extension is that we used a different definition of $\rho_\phi$ than in Refs.~\cite{Gomes:2013ema,Gomes:2015dhl}, mainly we kept the explicit dependence on $G_4$ in the left hand side of the Friedman equation \eqref{eq:1stfried}. In Refs.~\cite{Gomes:2013ema,Gomes:2015dhl}, the energy density of DE, say $\rho'_\phi$, is regarded as the remaining contribution after subtracting and adding $3H^2M_{pl}^2$ to Eq.~\eqref{eq:1stfried} so that it looks like $3H^2M_{pl}^2=\rho'_\phi+\rho_{DM}$. Obviously, physics do not depend on such choice of definition \cite{Bellini:2014fua,Bettoni:2015wla} but we easily miss solutions where $G_4$ plays an important role. The second reason for our generalization is that Refs.~\cite{Gomes:2013ema,Gomes:2015dhl} work in the matter frame and, therefore, the function $Q$ appears non-trivially in the master equation. Because of this one needs to use ansatz which need not be completely general. Thus, our new Lagrangian is more general that those previously found. 
        		
	\subsection{Phase space and stability of fixed points\label{subsec:phasespace}}
	
 Now that we have general DDE solutions for Horndeski model we will move to the analysis of their nature. In particular we will be interested in studying the phase space and see if the solutions found are attractors.   
    In order to do so, let us consider the Lagrangian given by Eq.~\eqref{eq:trackerlagrangian}. It is convenient to introduce the following variables:
		\ba\label{eq:definitions}
	x^2\equiv\frac{X\phi^{-2}}{3H^2}\quad{\rm ,}\quad y^2\equiv\frac{\phi^{p_2-p_4}}{3H^2}\quad{ ,} \quad \Omega_{DM}\equiv\frac{\rho_{DM}}{6H^2G_4}\quad ,\quad \Omega_b\equiv\frac{\rho_{b}}{6H^2G_4}\quad{\rm and}\quad \Omega_r\equiv\frac{\rho_{rad}}{6H^2G_4}\,.
	\ea
	Note that $Y=x^2/y^2$. In this way the first Friedman equation is given by
	\ba
	1&=\Omega_{\phi}+\Omega_{DM}+\Omega_b+\Omega_r
\quad{\rm where}\quad
	\Omega_{\phi}\equiv \frac{\rho_\phi}{6H^2G_4}\,.
	\ea
We can then write the second Friedman equation and the time derivative of the first Friedman equation respectively as
	\ba
	P_1\frac{dx}{dN}+P_2\frac{dy}{dN}+P=0 \quad{\rm and}\quad
	F_1\frac{dx}{dN}+F_2\frac{dy}{dN}+F=0
	\ea
	where
	\ba
	F=-(3+\sqrt{6}p_2 x)\Omega_{DM}-(3+\sqrt{6}\left(p_2-q\right) x)\Omega_b-(3+3w_{r}+\sqrt{6}p_2 x)\Omega_r
	\ea
	and the explicit expressions for $P$, $P_1$, $P_2$, $F_1$, $F_2$ and $\Omega_\phi$ can be found in the App.~\ref{app:formulas} due to their length. We have also used that
	\ba\label{eq:dm}
	\frac{d\Omega_{DM}}{dN}=-\Omega_{DM}\left(3+\sqrt{6}p_2 x+\frac{d\ln a_4}{dN}-2\frac{d\ln y}{dN}\right)\,,
	\ea
	\ba\label{eq:db}
	\frac{d\Omega_b}{dN}=-\Omega_b\left(3+\sqrt{6}\left(p_2-q\right) x+\frac{d\ln a_4}{dN}-2\frac{d\ln y}{dN}\right)\,,
	\ea
	\ba\label{eq:dr}
	\frac{d\Omega_r}{dN}=-\Omega_r\left(3(1+w_{r})+\sqrt{6}p_2 x+\frac{d\ln a_4}{dN}-2\frac{d\ln y}{dN}\right)\,,
	\ea
	where we made use of the fact that $\bar\rho_b\propto \bar a ^{-3}$ due to conservation of energy of baryons in the matter frame and that $\rho_b=\bar\rho_b B(\phi)^4$. The autonomous system of equations is given by Eqs.~\eqref{eq:dm}, \eqref{eq:db}, \eqref{eq:dr},
	\ba\label{eq:dxdn}
	\frac{dx}{dN}=\frac{1}{D}\left(F\,P_2-{P\,F_2}\right)\,
	\quad{\rm and}\quad
	\frac{dy}{dN}=\frac{1}{D}\left({P\,F_1-F\,P_1}\right)\,,
	\ea
	where
	$
	D=F_2\,P_1-F_1\,P_2
	$
    .
	The fixed point where $dx/dN=dy/dN=0$ is given by $F=P=0$.  Note that from the definition of $y$ we have
	\ba\label{eq:lny}
	2\frac{d\ln y}{dN}=\sqrt{6}\left(p_2-p_4\right)x+3(1+w_{\rm eff})\,.
	\ea
	The latter equation will be useful to relate $w _{\rm eff}$ with $p_2$ and $p_4$ at the fixed point. The general solution for $F=0$ is given by
	\ba\label{eq:xstot}
	x_{s}=-\sqrt{\frac{3}{2}}\frac{1}{p_2}\frac{\Omega_{DM}+\Omega_b+\Omega_r\left(1+w_{r}\right)}{\Omega_{DM}+\Omega_b\left(1-q/p_2\right)+\Omega_r}\,.
	\ea
	The equation $P=0$ will give us the solution for $y_s$. We can study if the solution is an attractor by looking at the perturbations around the solution $x=x_s+\delta x$ and $y=y_s+\delta y$. Denoting $\frac{\partial A}{\partial x}\equiv A_x$ we have that the perturbations are described by
		\ba
	\frac{d}{dN}\begin{pmatrix} \delta x \\ \delta y \end{pmatrix}=\frac{1}{D_s}
\hat{\cal M}
	\begin{pmatrix} \delta x \\ \delta y \end{pmatrix} \quad {\rm where}\quad\hat{\cal M}=
	\begin{pmatrix} -F_{2s}P_{x,s}+F_{x,s}P_{2s} & -F_{2s}P_{y,s}\\ F_{1s}P_{x,s}-F_{x,s}P_{1s} & F_{1s}P_{y,s} \end{pmatrix}
	\ea
	where a subindex $s$ indicates that the functions are evaluated on the fixed point solution. The eigenvalues of this matrix tells us how the perturbations grow or decay and are given by
	\ba\label{eq:mupm}
	\mu_{\pm}=\frac{{\rm Tr}\hat{\cal M}}{2D_s}\left(1\pm\sqrt{1-4\frac{\det\hat{\cal M}}{{\rm Tr}^2\hat{\cal M}}}\right)\,.
	\ea
	The system will be an attractor if $\mu_{\pm}<0$. The general form is involved and we shall use a particular example in next section.
	
	One may worry that an attractor in the dark matter frame might not be an attractor in the matter frame. This is clear once we take a look at the relation between variables. It can be checked that the variables of the autonomous system in the matter frame are
	\ba
	\bar x^2\equiv\frac{\bar X \phi^{-2}}{3\bar H^2} \quad{\rm and}\quad\bar y^2\equiv\frac{\phi^{p_2-p_4-2q}}{3\bar H^2} \,.
	\ea
	The relation with the dark matter frame variables is given by
	\ba
	x=\frac{\bar x}{1-\sqrt{6}q\bar x} \quad{\rm and}\quad y=\frac{\bar y}{1-\sqrt{6}q\bar x}\,.
	\ea
	It is clear from this that the attractor behavior is not substantially changed. Perturbing around the DDE solution with constant $x$ and $y$ just gives a constant rescaling relating $x$ and $\bar x$. Regarding $y$, it mixes $\bar y$ with $\bar x$ but this will not change the attractor behavior. The relevant change would be that 
	\ba
	\delta x \propto a^{\mu_{\pm}} \quad{\rm whereas} \quad \delta \bar x \propto \bar a^{\bar \mu_{\pm}} 
	\ea
	where we just used that $a = B^{-1} \bar a$ and then
	\ba\label{eq:barmupm}
	\bar \mu_{\pm}={{\mu_\pm}}\left(1-\beta\right) \quad {\rm with}\quad \beta=\frac{\sqrt{6}qx}{1+\sqrt{6}qx}\,.
	\ea
    At this point note that $\beta$ is a free parameter (given by the free parameter $q$), only appearing through the relations of $c$, $w_{\rm eff}$ with $\bar c$ and $\bar w_{\rm eff}$. It is interesting to see that a priori by choosing $\beta<0$ and large we can make our solution a very strong attractor. We will see however that it cannot be made an infinitely strong attractor due to the implicit dependence on $\beta$ in $c$. Also note how for $\beta>1$ the solution is apparently no longer an attractor in the matter frame. To understand this take a look at the relation between the number of e-folds $dN=\left(1-\beta\right)d\bar N$. Take for example $\beta=-9$. It means that $10$ e-folds in the dark matter frame corresponds to $1$ e-fold in the matter frame. Thus, the attractor is reached in less e-folds using the matter frame time coordinate. Contrariwise, if $\beta>1$ the direction of time in the matter frame is reversed and thus the system is getting out of the attractor as time goes.
	
	\section{Applications to {DE}: attractors with $\mathbf{c_{T}=1}$\label{sec:aplications}}
	
	Let us apply our newly derived model as a viable DE model. We need our model to be compatible with $c_{T}^2/c^2-1<10^{-15}$. A study of the tensor perturbations of our models in the fixed point yields (see App.~\ref{app:perturbations})
	\ba
	c_T^{2}=\frac{a_4-p_5 Ya_5}{a_4-2Ya_{4,Y}+p_5 Ya_5-\left(6+p_2-3p_4\right)Y^2a_{5,Y}}\,.
	\ea
	In order to satisfy the LIGO constraint one possibility is to take $a_{4}=M_{pl}^2/2$ and $a_5=0$, i.e., to reduce to KGB model \cite{Deffayet:2010qz}. Note that $p_4$ (the exponent of the non-minimal coupling) does not appear in $c_T^2$, as any conformal coupling that depends only on $\phi$ does not modify the propagation of GWs. The second possibility is that $a_4$ and $a_5$ are such that their combination in $c_T^2$ cancels out. However, we note that this would require an a priori unjustified fine-tuning \cite{Ezquiaga:2017ekz,Creminelli:2017sry,Baker:2017hug}. A third option is to extend the discussion to beyond Horndeski and extended scalar tensor theories (a.k.a. DHOST) and select those models where \textit{at linear level} $c_T=1$ \cite{Crisostomi:2017pjs,Langlois:2017dyl}. 
    
    Here, we will instead investigate a fourth possibility which is characteristic of DDE solutions. To satisfy the constraint we require that
	\ba
	a_{4,Y}\big|_{s}=0 \quad {\rm ,}\quad a_{5,Y}\big|_s=0\,,
	\ea 
	and $a_{5}\big|_s=0$ or $p_5=0$ \textit{evaluated on the DDE} solutions. In other words, on the DDE solution $G_4$ should effectively depend on $\phi$ {only} and $G_5$ should be constant. The interesting feature of this mechanism is that this requirement would be dynamically reached (as long as $\mu_\pm<0$) and only applies when we are on the DDE solution and therefore out of such solution $a_{4,Y}\neq0$ and $a_{5,Y}\neq0$ in general. This is is then able pass the LIGO constraint because the detection is at $z\sim 0.08$ which means that occurred in our nearby universe (in cosmological terms).
    
    At this point, however, one has to be sure that $c_T^2=1$ is actually stable. Let's consider a small perturbation out from the DDE solution, e.g. due to the effect of baryons. Then for $Y=Y_s+\delta Y$ we have
	\ba
	\delta c_T^{2}=\delta Y\frac{2Y_s a_{4,YY}|_s+\left(6+p_2-3p_4\right)Y_s^2a_{5,YY}|_s}{a_4|_s}\,.
	\ea
	Since the constraint from observations is extremely tight we shall require as well $a_{4,YY}|_s=a_{5,YY}|_s=0$. In turn this will simplify considerably the equations on the DDE. In this way, only non-linear effects will cause a departure from $c_T|_s=1$, for example during radiation domination. In what follows we will consider the case where $a_3=a_5=0$ and $a_{4,Y}|_{s}=a_{4,YY}|_{s}=0$.
	
	Before proceeding further let us check the no-ghost condition for the case where $a_3=a_5=0$ and $a_{4,Y}|_{s}=a_{4,YY}|_{s}=0$. We will use the formulas derived in Ref.~\cite{DeFelice:2011bh} and for completeness we wrote them in the App.~\ref{app:perturbations}. The no-ghost conditions for the gradient and kinetic terms of the perturbations respectively are $c^2_s>0$ and $Q_s>0$ which read
	\ba\label{eq:cs}
	c_s^2\propto(2p_4-p_2) \left(3 p_4 x_s+\sqrt{6}\right)x_s-3 \Omega_{DM}-3\Omega_b-3 \Omega_r(1+w_{\rm rad})>0
	\ea
	and
	\ba
	Q_s&\propto x_s\left(2p_4-p_2\right)\left(\sqrt{6}+3p_4x_s\right)-3\Omega_b-3 \Omega_{DM}-3 \Omega_r+\frac{x_s^2a_{2,YY}}{a_4}
	\\&+\frac{12Y_s^3a_{4,YYY}}{a_4}\left(2+\sqrt{6}\left(2+p_2-p_4\right)x_s\right)>0\,,
	\ea
	where we used $P=0$ to solve for $a_2$ and the first Friedman equation \eqref{eq:1stfried} to solve for $a_{2,Y}$. The condition $Q_s>0$ is easily achieved if $c_s^2>0$, $a_{2,YY}>0$ and $a_{4,YYY}/a_4\ll1$ or if the last term is positive. Let us study the case when DDE dominates ($\Omega_r\to0$) and when radiation dominates ($\Omega_{DM},\Omega_b\to0$). Neglecting the effect of baryons, we respectively find that $c_s^2=0$ has two solutions on the DDE fixed point, namely
	\ba\label{eq:p4p2}
	\frac{p_{4\pm}}{p_2}=\frac{1}{12}\left(7\pm\sqrt{1+48 \Omega_{DM}}\right)
\quad{\rm and}\quad
\frac{p_{4\pm}}{p_2}=\frac{1}{2}\left(1\pm\sqrt{ \Omega_r}\right)\,,
	\ea
	where we already used Eq.~\eqref{eq:xstot}. A short calculation shows that    $c_s^2>0$ if $p_4\notin\left(p_{4-},p_{4+}\right)$. This also implies that for $q=0$, i.e. $\Omega_{DM}$ and $\Omega_r$ are independent of DE, the solution $p_4=0$ is always safe since in that case ${p_{4\pm}}/{p_2}>0$, as it should be. We can rewrite the condition Eq.~\eqref{eq:cs} on the DDE solution in terms of $\bar w_{\rm eff}$, $\bar c$ and $\beta$ using Eqs.~\eqref{eq:ctobarc} and \eqref{eq:weff2}. The positivity of $c_s^2$ can be then translated to the fact that $\beta\notin(\beta_-,\beta_+)$, where
	\ba\label{eq:betapm}
	\beta_{\pm}=\frac{1}{2}\left(5+9\bar w_{\rm eff}\right)\pm \frac{1}{2}\sqrt{1+3\bar w_{\rm eff} \left(2+3\bar w_{\rm eff}\right)+\frac{24}{1+\bar c}}\,.
	\ea
	For example, for $\bar c=2.3$ and $\bar w_{\rm eff}=-0.7$ we have that $\beta\notin(-2.1,0.8)$. Outside this range the theory is healthy. It should be noted that
	if we consider the case $\bar w_{\rm eff}=-1$ then $\beta\notin(-3.7,-0.33)$ and the original model with $p_4=p_2/2$ has a ghost in general.
    
    There is an interesting result from our analysis. Only considering $G_2$ and $G_4$ (which includes the original models), we have found that we need to consider a non-universal coupling to gravity if we require Doppelg{\"a}nger behaviour, acceleration and stable perturbations. The reason is that if DM and SM universally couple to gravity, that is $\beta=0$ (matter and dark matter frames coincide), and we require that $\bar w_{\rm eff}=w_{\rm eff}=-p_4/p_2\sim-0.7$ we find that there is a ghost in general, i.e. $c_s^2<0$. 
    We could consider a more general Lagrangian with a suitable $G_3$ but rather than entering in more fine-tunings we will stick to the non-universal coupling to gravity. For this reason, we will consider the case where $\beta<\beta_-$ which is both healthy and interesting as we shall see. The case $\beta\sim\beta_{-}$ is also interesting for models where DE could cluster as $c_s^2\sim0$.

	\subsection{Phase space and stability of fixed points of DDE\label{subsec:stability}}
	
	Let us now study in detail the phase space and stability of this particular example. We will apply the equations derived in Sec.~ \ref{subsec:phasespace} and App.~\ref{app:formulas} for $b=r=0$ and $a_{4,Y}=a_{4,YY}=a_{4,YYY}=0$. The latter equality will be justified a posteriori. Note that this case is a non-minimally coupled quintessence and generalizes previous results in Ref.~\cite{Amendola:1999qq,Tsujikawa:2004dp} and reduces to them when $\bar p_4\to 0$. The Friedman equation is now $1=\Omega_\phi+\Omega_{DM}$ where
	\ba\label{eq:omega}
	\Omega_{\phi}={x^2} \frac{a_{2,Y}}{a_4}-y^2\frac{a_{2}}{2a_4}-\sqrt{6}p_4x=\frac{c}{1+c}	\,,
	\ea
	and we used the DDE condition $\rho_\phi/\rho_{DM}=\Omega_{\phi}/\Omega_{DM}=c$. We also have
	\ba
	F=-(3+\sqrt{6}p_2 x)\Omega_{DM}=0 \qquad\Rightarrow\qquad x_s=-\sqrt{\frac{3}{2}}\frac{1}{p_2}\,.
	\ea
	From Eq.~\eqref{eq:lny} we find
	\ba
	w_{\rm eff}=-\frac{p_4}{p_2}\,.
	\ea
	The remaining condition is given by
	\ba
	P=1+\frac{1}{3}x\left(p_2+{p_4}{}\right)\left(\sqrt{6}+3p_4 x\right)+\frac{y^2a_2}{2a_4}=0\,,
	\ea
	which we will use to solve for $a_2$. We study the perturbations around the fixed point and in this particular case Eq.~\eqref{eq:mupm} yields (see App.~\ref{app:formulas})
	\ba\label{eq:mupm}
	\mu_{\pm}=-\frac{3}{4}\left(1-\frac{p_4}{p_2}\right)\left\{1\pm\sqrt{1-8\frac{1-\Omega_{\phi}}{A\left(1-{p_4}/{p_2}\right)^2}\left(2\Omega_\phi+\frac{p_4}{p_2}\left(3\frac{p_4}{p_2}-5\right)\right)}\right\}
	\ea
	where
	\ba
	A\equiv2\Omega_\phi+\frac{p_4}{p_2}\left(6\frac{p_4}{p_2}-7\right)+\frac{9}{p_2^4y^2}\frac{a_{2,YY}}{a_4}\,.
	\ea
    Note that $A\propto Q_s$ and therefore the no-ghost condition imposes $A>0$ as well. A short exercise tells that Eq.~\eqref{eq:barmupm} applied to the first example of Sec.~\ref{subsec:comparison} exactly matches the results of Ref.~\cite{Tsujikawa:2006mw}. Thus, it is a further support of our calculations in the dark matter frame.

Let us consider the first non-trivial extension of Ref.~\cite{Amendola:1999qq,Tsujikawa:2004dp}, that is canonical scalar field ($a_{2,YY}=0$) with a general non-minimal coupling to gravity ($p_4\neq 0$). We are interested in the case where the system is a strong attractor in the matter frame, i.e., $\bar\mu_{\pm}<0$. According to Eq.~\eqref{eq:barmupm}, we may choose $\beta$ very large so as to have $|\bar\mu_{\pm}|\gg1$. However, a quick inspection to Eqs.~\eqref{eq:weff2}, \eqref{eq:ctobarc} and \eqref{eq:omega} tells us that in the limit $\beta\to -\infty$ we are led to $p_4/p_2\to1/3$ and $\Omega_\phi\to 1$. In that limit, $\bar\mu_-\to0$ as the last term of the square root in Eq.~\eqref{eq:mupm} goes to zero as $\beta^{-2}$. Contrariwise, if $\beta=\beta_-\sim -2.1$ (the upper bound for the no-ghost conditions) we find that the square root becomes imaginary and thus $\bar\mu_{-}\approx-1.3$. A numerical search finds that the optimal value is $\beta\approx-3.8$ where $\bar\mu_-\approx -2.2$. This means that in $1$ e-fold the system approaches the attractor by $0.1$. For a general form of $a_{2,YY}\neq0$ one could make the attractor much stronger. In any case, the main point of this section is to show that the simplest model is generally an attractor. For this reason, we expect that by enlarging the functional space to include $a_3$ and $a_5$ there will still be models with such attractor behavior. We will see that in fact the main issue with this model will be a departure from the DDE due to baryons.

	\subsection{Effect of baryons}
 In the DM frame we have seen how baryons get non-minimally coupled to the scalar field. So far we have neglected this component as it is subdominant in the late time cosmology. However, the effect of baryons is quite interesting. First of all, it is important to note that on the attractor solution
	\ba
	\Omega_b=\frac{\rho_b}{6H^2G_4}\propto a^{-\frac{3q}{p_2}}=a^{\frac{\beta}{1-\beta}}\,,
	\ea
	where we integrated Eq.~\eqref{eq:db} on the fixed point.
	For $\beta<0$  or $\beta>1$ ($q/p_2>0$) we have that the relative energy density of baryons increases backwards in time. This means that there was an epoch where baryons  dominated the universe. If $0<\beta<1$ ($q/p_2>0$) then $\Omega_b$ increases with time and baryons will dominate in the future. For $\beta=q=0$ baryons interact with DE like DM and the DDE solution is preserved but this case has a ghost in the scalar sector (see Sec.~\ref{subsec:stability}). We will treat baryons perturbatively and study its effects. The effect of baryons into the scaling value of $c\to c+\delta c$ is small and at the current time is given by
	
	\ba
	\frac{\delta c}{c}=\left(1-\frac{\bar c}{c}\right)\frac{\bar\Omega_{b,0}}{1-\bar\Omega_{b,0}}=\frac{1+{c}^{-1}}{\left(1-\beta\right)^2}\frac{\bar\Omega_{b,0}}{1-\bar\Omega_{b,0}}\approx \frac{4\times10^{-2}}{\left(1-\beta\right)^2} \,
	\ea
	where we took $\bar\Omega_{b,0}\equiv{\bar\rho_{b,0}}/{3\bar H_0^2M_{pl}^2}\approx4\times10^{-2}$ in the matter frame and that $c>1$ since $|\beta|>1$ ($\beta<0$), see Eq.~\eqref{eq:ctobarc}. Solving $\delta P=\delta F=0$ due to baryons we find
	\ba\label{eq:changexy}
	\frac{\delta x}{x_s}=\frac{q}{p_2}\frac{\bar \Omega_{b}/\bar\Omega_{DM}}{1+\left(1-q/p_2\right)\bar\Omega_{b}/\bar\Omega_{DM}}
	\quad {\rm and}\quad
	\frac{\delta y}{y_s}=\frac{\delta x}{x_s}\left(1-\frac{1-p_4/p_2}{\Omega_{\phi}+\frac{p_4}{2p_2}\left(3\frac{p_4}{p_2}-5\right)}\right)\,,
	\ea
	where we have used $\bar\Omega_{DM,0}\equiv{\bar\rho_{DM,0}}/{3\bar H_0^2M_{pl}^2}\approx 0.27$ and that the ratio $\bar\Omega_{b}/\bar\Omega_{DM}=\Omega_{b}/\Omega_{DM}$ is frame independent. Note that for $\beta,q\to 0$ there is no effect from the baryons.
	We can now compute the change in $Y_s=x_s^2/y_s^2$ as
	\ba
	\frac{\delta Y}{Y_s}=2\frac{\delta x}{x_s}\frac{1-p_4/p_2}{\Omega_{\phi}+\frac{p_4}{2p_2}\left(3\frac{p_4}{p_2}-5\right)}\,.
	\ea
	The effect on $c_T^2$, if we assume that the previous $(n-1)$-th derivative of $a_4$ vanish, i.e. $a_{4,Y^{n-1}}|_s=0$, reads
	\ba
	\delta c_T^{2}=\frac{\delta Y^{n-1}}{Y_s^{n-1}}\frac{2Y^{n}_s a_{4,Y^{n}}|_s}{a_4|_s}\,.
	\ea
	To have an order of magnitude estimate let us use that $\Omega_{b}/\Omega_{DM}\approx 0.15$ and assume that $\beta<\beta_{-}\sim-2.1$. In that case, we find that typically $\delta Y/Y_s\sim 0.1$ (since for large $\beta$ we have $q/p_2\sim p_4/p_2 \sim 1/3$ and $\Omega_\phi\sim 1$) and we can roughly estimate
	\ba
	\delta c_T^{2}\approx\,10^{-n+1}\frac{2Y^{n}_s a_{4,Y^{n}}|_s}{a_4|_s}\,.
	\ea
	Let us assume that $a_4$ has a ``minimum'' in Y, e.g.
	\ba\label{eq:a4}
	a_4(Y)=\frac{M_{pl}^2}{2}\left(1+c_{4}\left(1-\frac{Y}{Y_s}\right)^n\right)\,,
	\ea
	where $n$ represent the steepness and the typical value of $Y_s\propto H_0^2$ gives us the scale in which the DDE with $c_T=1$ starts. The constraint from GWs then tells us that
	\ba
	\delta c_T^{2}\approx\,-n\,10^{-n+1}c_{4}<10^{-15}\,.
	\ea
	For example, if we require that $c_{4}\sim O(1)$ we need $n>16$. As expected we need a large tuning to be compatible with such a tight constraint. We may relax the value of $n$ by assuming that $c_{4}\ll 1$ but then one may argue that we are fine tuning the coefficient as well. For completeness, we check the effect on the effective equation of state which is small, as expected, and it is given by
	\ba
	\delta w_{\rm eff}=(1+w_{\rm eff})\frac{\delta x}{x_s}\approx 10^{-2}\,.
	\ea
    
    Let us briefly discuss possible screenings on local scales. Since we have both a conformal coupling to baryons and higher derivatives the model potentially has Vainshtein and Chameleon screenings. The length scale of the Vainshtein mechanism is given by the coefficient in front of the higher order derivatives \cite{Kimura:2011dc,Babichev:2013usa}, that is $G_{4,X}$. Since we are expanding around a cosmological background where $G_{4,X}$ is vanishing we have that the Vainshtein will be essentially zero for practical purposes. It should be noted that by considering a non-trivial $G_3$, we could enlarge the possibility of Vainshtein screening. 
    
    Let us turn now to the Chameleon screening. Since in the matter frame we have a conformal coupling to baryons with $q$, the Chameleon mechanism \cite{Khoury:2003rn,Khoury:2003aq} would apply for baryons depending on the effective potential for $\phi$. For example, for simplicity we can consider that the Lagrangian for baryons is similar to that of DM, i.e. Eq.~\eqref{eq:DM}, but for the metric $\bar g$. It is convenient to go to the ``Einstein'' frame\footnote{We regard the ``Einstein'' frame by the frame where $\tilde G_4\propto \tilde a_4(Y)$, that is a constant factor on the DDE solution.} by $\bar g_{\mu\nu}=\phi^{-\bar p_4} \tilde g_{\mu\nu}$. In such frame, we can see that the effective potential is roughly given by (if $\bar p_4\neq0$)
	\ba
	V_{\rm eff}=
    \phi^{-2 p_4}\left(V_0\phi^{p_2}+\frac{\bar \rho_b}{4}\phi^{4q}\right)\,.
	\ea
	There will be possibility of screening where the baryon energy density is relevant if \ba \left({p_2}/{p_4}-2\right)  \left( {q}/{p_4}-{1}/{2}\right) < 0\ea
	For our particular case ($\beta<0$ and $|\beta|\ll1$), we have $p_4/p_2\sim1/3$ and $q/p_2\sim1/3$ which does not fall in the Chameleon screening. In fact only for $3\bar w_{\rm eff}<\beta<3(1 + 2 \bar w_{\rm eff})$ will there be screening mechanism. It is interesting to note that it falls in the excluded regime by the no-ghost conditions. The only way out is to consider that $\bar p_4=0$ (alternatively $\beta\sim -2.1$) when the matter frame is already the ``Einstein'' frame. There will not be any screening but there will not be any fifth force either, like quintessence models \cite{Tsujikawa:2004dp}. A further study might be interesting but it is out of the scope of the present work. Here we present the minimal example where $c_T^2=1$ is not achieved by a fine tuning of the coefficient but rather by the presence of a ``minimum'' at the present time for the function $G_4$.
    
    We end this section by suggesting possible ways to attain a proper matter domination and radiation stages and to study the modification of $c_T$ at early times. The first point to note is that the solutions on the dark matter frame do not depend directly on $\beta$ (the coupling to the SM). They do depend indirectly once we require that baryons see an accelerated expanding universe today with $\bar w_{\rm eff}\approx -0.7$. An interesting possibility is to allow for a time dependence in $\beta$ -- essentially constant nowadays but changed in the past. Then we can see from Eqs.~\eqref{eq:weff2} and \eqref{eq:ctobarc} that $\bar w_{\rm eff}$ and $\bar c$ are not constant and they could, for example, change towards a matter dominated stage. This also implies that there must be DDE solutions without constant effective equation of state. Regarding the value of $c_T^2$ during radiation domination, we would require a specific model that has a proper matter domination stage to study the full evolution. For example, significant departures from $c_T=1$ during matter and radiation domination eras could be seen respectively by space-based GWs detectors like LISA or by B-mode polarization ($c_T^2\neq1$ shifts the positions of the angular peaks in the power spectrum, as first pointed out in \cite{Amendola:2014wma}). For example, if $r\sim 0.1$ the CMB bound is $c_T^2<3$ \cite{Raveri:2014eea}. For smaller $r$, the constraint is looser since $r\propto c_T^{-1}$ (with all the other parameters fixed). Nevertheless, we can give a rough estimates on the deviation from $c_T^2=1$ studying the change in $Y$. From its definition (see Eq.~\eqref{eq:definitions}) we see that on the fixed point $Y\propto H^2\phi^{p_4-p_2-2}={\rm cnt}$. To compare the values of $Y$ during radiation and DDE we need to know the evolution of $\phi$ as well -- which will try to track that of $H$. During radiation domination we can see that $x_r=x_s\left(1+w_{r}\right)$ (by see Eq.~\eqref{eq:xstot}). Also note that a similar calculation than in Eq.~\eqref{eq:changexy} but for radiation instead of baryons, yields that $\delta y/y_s<0$. While $x$ grows, $y$ decreases. We can thus place a lower bound to the value of $Y=x^2/y^2$ during radiation domination, namely $Y_{r}>Y_s\left(1+w_r\right)^2$ ($|1-Y_{r}/Y_{s}|> 7/9$). Eq.~\eqref{eq:a4} implies that for large $n$ the deviations from $c_T^2=1$ could be significant, namely
\ba
    c_{T,r}^2&=\frac{1+c_4\left(1-Y_r/Y_s\right)^n}{1+c_4\left(1+\left(2n-1\right)Y_r/Y_s\right)\left(1-Y_r/Y_s\right)^{n-1}}\,.
\ea
We will get a lower or upper bound on $c_T^2$ depending on the values of $c_4$ and $n$. For example, for $c_4\sim -1$ and $n=16$ we have that $c_T^2<0.44$ while for $c_4\sim 0.1$ we get $c_T^2>1.15$. Lastly, we note that for $c_4>0.8$ we would have a ghost, i.e. $c_T^2<0$. Similar logic applies to odd $n$ by flipping the sign of $c_4$. The main point is that given a complete model for DDE cosmology we would be able to constraint the value of $c_4$ and $n$ using the early/late time universe bounds on $c_T^2$ \cite{LIGO:2017qsa,Amendola:2014wma,Pettorino:2014bka,Raveri:2014eea}. Future observations of CMB B-mode polarization might provide constraints on the parameters. For the LISA band, it will depend on how to achieve the matter dominated stage. For example, if it is achieved by a time-dependent $\bar w_{eff}$ we do not expect much deviation form $c_T^2=1$ since $x$ will be roughly constant. Thus, we have provided a model where significant deviations from $c_T^2=1$ in the early universe are expected.

\section{Conclusions\label{sec:conclusions}}
	
The (almost) simultaneous detection of GWs and their electromagnetic counterpart \cite{LIGO:2017qsa} ruled out, at first glance, most of the Horndeski theories \cite{Ezquiaga:2017ekz,Creminelli:2017sry,Baker:2017hug}; basically all the terms that contain derivative couplings to gravity, i.e. ${\cal L}_4$ and ${\cal L}_5$. Here we enlarged the space of models that could potentially pass the GW constraint within Horndeski theories \emph{with interaction between the scalar field and dark matter}. We proposed a class of models with non-trivial ${\cal L}_4$ and ${\cal L}_5$ in which the value $c_T=1$ might be achieved dynamically and, therefore, avoids the fine-tuning problem. For simplicity, we studied the particular cases without ${\cal L}_3$ and ${\cal L}_5$ and show that there are attractor solutions with $c_T=1$. We expect that there are still solutions including ${\cal L}_3$ and ${\cal L}_5$ with attractor behavior, since the functional space has been enlarged. Furthermore, these models can be take as a motivation to consider effective field theory models of dark energy \cite{Gubitosi:2012hu} in which $c_T(t)\to1$ only at low redshifts but $c_T\neq1$ at high redshifts.
   
To do that, we found new solutions to interacting dark sector models in which the ratio between dark energy and dark matter energy densities is constant; in turn alleviating the coincidence problem. We called these class of solutions Doppelg{\"a}nger Dark Energy (DDE).
DDE models are usually interpreted as a non-trivial interaction between DM and DE. 
In this work, we have provided a new interpretation of the model by removing the interaction via a conformal transformation. We then introduced the \textit{matter frame} where baryons are minimally coupled but DM interacts with DE and the \textit{dark matter frame} where the DM is a free dust fluid but baryons have a dilatonic coupling to DE. In the latter frame DDE solutions are viewed just as regular DM plus a DE component which behaves like a matter fluid (in the DDE regime). The observed accelerated expansion of the universe is then due to a conformal coupling between DE and the standard model. We have found the most general solutions of DDE in the Horndeski Lagrangian; thus greatly extending the results in the literature \cite{Gomes:2015dhl}. One of the main results is the general form of the Lagrangian which admits DDE solutions and it is given in the matter frame by Eq.~\eqref{eq:matterframelagrangian}.
	
	Concerning the GW bounds on $c_T$, we discussed the theory space that is still allowed that includes DDE solutions with a non-trivial form of ${\cal L}_4$ and ${\cal L}_5$. The crucial point is that ${\cal L}_i$ are general functions of $Y\equiv X\phi^{2p}$ which is constant on the DDE solution. In this way, we chose that $G_4$ and $G_5$ to have a ``minimum'' in $Y$ only on the attractor solution, say $G_{4,Y}|_s=G_{5,Y}|_s=0$ and therefore $c_T^2=1$, but not otherwise. Afterwards, we focused on a particular model within the new solutions using only ${\cal L}_2$ and ${\cal L}_4$. Interestingly, these DDE solutions are attractors for a certain parameter range; thus, reaching the value $c_T=1$ at low redshifts dynamically.     
	
    We have then studied the phase space of the system and we have imposed the no-ghost conditions for perturbations. We found that the no-ghost conditions on accelerating DDE solutions with general $G_2$ and $G_4$ require a non-universal coupling to gravity. Assuming the $\Lambda CDM$ values, that is $\bar w_{\rm eff}\approx-0.7$ and $\bar c\approx2.3$ (see Eqs.~\eqref{eq:weff2} and \eqref{eq:ctobarc}), we have that the value of the dark matter frame variables $c$ and $w_{\rm eff}$ depend only on the conformal coupling to matter $\beta$.
    The model is stable and an attractor for $\beta<\beta_{-}$, which is clear from Eqs.~\eqref{eq:betapm} and \eqref{eq:mupm}. Furthermore,  we estimated the steepness of the ``minimum'' in $G_4$ by a parameter $n$ (${\partial^i G_4}/{\partial Y^i}=0$ for $i<n$) which tells us how hard it is to depart from $c_T^2=1$ with a departure from the DDE solution. We have found that due to the effect of baryons, which in general takes the system out of the fixed point by $1\%$, the power $n$ has to be fairly large. In fact, the repercussion of the effect to the departure of $c_T^2$ scales as $\delta c_T^2\sim n\,\left(\Omega_b/\Omega_{DM}\right)^{n-1}$ and thus we must require that $n>16$ in order to be compatible with the observation of the GW event \cite{LIGO:2017qsa}. We have argued that our model predicts significant departures from $c_T^2=1$ during radiation domination, which might place future bounds on our parameters using CMB B-mode polarization data. Future observations on CMB B-mode polarization \cite{Amendola:2014wma,Pettorino:2014bka} and space-based GWs detectors, e.g. LISA, will place stringent constraints on this kind of models. A glance at possible screening mechanisms shows that neither the Chameleon nor the Vainshtein mechanisms would not work in our particular model. Nevertheless, while this particular example might be ultimately ruled out by other constraints, we have proposed a dynamical mechanism to achieve $c_T^2=1$ that goes beyond those discussed in Refs.~\cite{Ezquiaga:2017ekz,Creminelli:2017sry,Baker:2017hug,Langlois:2017dyl,Crisostomi:2017pjs,Sakstein:2017xjx}; yet it allows for significant departures from $c_T^2=1$ in the early universe. Although the presented models require a fine-tuning of $n>16$ and, thus, they might not be distinguishable from other forms of tuning, they dynamically achieve $c_T^2\to 1$ at present only with $G_4$ and are essentially not background dependent.

	Let us end by noting that our approach could also be applied to generalized interactions and generalized models. For example, we could have a DDE in the dark matter frame and consider a general disformal coupling to matter with kinetic dependence. Then we should require that in the matter frame $c_T^2=1$ but we will have a non-trivial derivative interaction between DM and DE. This line of research will be pursued elsewhere. It would also be interesting to derive non-universal couplings to gravity in the dark sector and in the standard model from a fundamental approach but this is far from the scope of this paper. 
	
\begin{acknowledgments}
	G.D. would like to thank A. de Felice, A. Naruko, J. Rubio, R. Saito and J. Takeda for useful discussions. L.A. and D.B. acknowledge financial support from the SFB-Transregio TRR33 ``The Dark Universe''. G.D. acknowledges the support from DFG Collaborative Research centre 
SFB 1225 (ISOQUANT). A.R.G. thanks CNPq and FAPEMA for financial support. G.D. also thanks the Yukawa Institute for Theoretical Physics at Kyoto University. Discussions during the YITP symposium YKIS2018a "General Relativity -- The Next Generation --" were useful to complete this work.
	\end{acknowledgments}
	
	\appendix
	
	\section{Mapping frame to frame\label{app:rules}}
	Let us compute the redshift in the dark matter frame. As it is well explained in Ref.~\cite{Deruelle:2010ht} if the SM has a non-trivial coupling to a scalar field one finds that the mass of the baryons and fermions are rescaled under Eq.~\eqref{eq:DMtoSM} by $m=B \bar m$ so that our knowledge of emission of photons has to be translated into the past. For example, when we compare a observed frequency from a transition at a time $t$ and today we find
	\ba
	\nu(t)=\frac{B(t)}{B_{0}}\nu_{0}\,,
	\ea
	where the subindex $0$ stands for today. Thus, we one computes the redshift it does not only contain information about the expansion but about the time dependent mass of the particles as well. The redshift of the photons then can be written as
	\ba
	1+z=\frac{\nu_{em,0}}{\nu_{obs,0}}=\frac{B_{obs,0}}{B_{em}}\frac{\nu_{em}}{\nu_{obs,0}}=\frac{B_{obs,0}}{B_{em}}\frac{a_{em}}{a_{obs,0}}=\frac{\bar a_{em}}{\bar a_{obs,0}}\,,
	\ea
	where in the first step we are measuring the frequency as it would be emitted today and has to be translated to the corresponding time of emission to related it with the expansion of the universe. Note that this coincides with the usual calculation in the matter frame.
	
	A similar reasoning can be done for the distance luminosity relation and one finds
	\ba
	D_{L}=\frac{\nu_{em,0}}{\nu_{obs,0}}r=\left(1+z\right)\int\frac{dt}{a}=\left(1+z\right)\int\frac{dz}{a}\frac{dt}{dz}=\left(1+z\right)\int dz\frac{B}{H\left(1+\frac{d\ln B}{dN}\right)}=\left(1+z\right)\int\frac{dz}{\bar H}
	\ea
where $r=\int dt /a $ is the physical distance travelled by the photons. Thus, observables are frame independent as it is well known.

For example, in $\Lambda CDM$ we have that $B=1$, $Q=0$ (DM behave as a pressureless fluid) and thus
	\ba
	3\bar H^2=\bar\rho_{\Lambda}+\bar\rho_{DM,0} \bar a^{-3}\quad {\rm and}\quad 1+\bar w_{\rm eff}=\frac{\bar\rho_{DM,0} \bar a^{-3}}{\bar\rho_{\Lambda}+\bar\rho_{DM,0} \bar a^{-3}}\,,
	\ea
	where a subindex $0$ refers to the value today and $a_0=1$. Then we conclude that at present $\bar w_{\rm eff}\approx-0.7$ which yields that $\bar \rho_{\Lambda}/\bar \rho_{DM}\approx 2.33$.

\subsection{Change in the Lagrangian}
Here we derived the relations between Lagrangian up to $G_4$. If the reader is interested in $G_5$, it is derived in Refs~\cite{Bettoni:2013diz}. Now, given that 
\ba
\bar g_{\mu\nu}= B(\phi)^2 g_{\mu\nu}
\ea
we find
\ba
\bar\n_\mu\bar\n_\nu\phi=\n_\mu\n_\nu\phi-2\frac{B_\phi}{B}\left(\n_\mu\phi\n_\nu\phi+Xg_{\mu\nu}\right)
\quad{\rm and}\quad
\bar R= B^{-2}\left(R-6\frac{\Box B}{B}\right)\,.
\ea
Using this relations it is straightforward to show that
\ba
G_4=B^2\bar G_4\quad,\quad G_3=B^2 \bar G_3+4 {B_\phi}{B} X \bar G_{4,X}+2\bar G_{4}{B_\phi}{B}
\ea
and
\ba
G_2 = B^4 \bar G_2+4B B_\phi \bar G_3 X+4X \bar G_4 B B_{\phi\phi}-8\bar G_4 B_{\phi}^2 X-8{B_\phi}{B}\bar G_{4,\phi} X\,,
\ea
where the arguments of the barred functions are now to be intended as functions of the matter frame variables, i.e. $\bar X= B^{-2}X$.

	\section{Explicit formulas \label{app:formulas}}
	Here we present for completeness the form of Horndeski equations of motion terms. They are given by
	\ba
	{\cal E}_2&=2XG_{2,X}-G_2\quad,\quad
	{\cal E}_3= 6X\dot\phi HG_{3,X}-2XG_{3,\phi}\,,\\
	{\cal E}_4&=-6H^2G_4+24H^2X\left(G_{4,X}+XG_{4,XX}\right)-12HX\dot\phi G_{4,\phi X}-6H\dot\phi G_{4,\phi}\,,\\
	{\cal E}_5&=2H^3X\dot\phi\left(5G_{5,X}+2XG_{5,XX}\right)-6H^2X\left(3G_{5,\phi}+2XG_{5,\phi X}\right)\,
	\ea
	and
	\ba
	{\cal P}_2&=G_2\quad,\quad
	{\cal P}_3= -2X\left(G_{3,\phi}+\ddot\phi G_{3,X}\right)\,,\\
	{\cal P}_4&=2\left(3H^2+2\dot H\right)G_4
	-\left(4X\left(3H^2+2\dot H\right)+4H\dot X\right)G_{4,X}
	-8HX\dot X G_{4,XX}\\&+2\left(\ddot\phi+2H\dot\phi\right)G_{4,\phi}+4XG_{4,\phi\phi}+4X\left(\ddot\phi-2H\dot\phi\right)G_{4,\phi X}\,,\\
	{\cal P}_5&=-2X\left(2H^3\dot\phi+2H\dot H\dot\phi+3H^2\ddot\phi\right)G_{5,X}-4H^2X^2\ddot\phi G_{5,XX}\\&+4HX\left(\dot X- HX\right)G_{5,\phi X}+2\left(2\dot H X+2\dot X H +3H^2X\right)G_{5,\phi}+4HX\dot\phi G_{5,\phi\phi}\,.
	\ea
	The formulas for the general phase space are given by
	\ba
	1=\Omega_\phi+\Omega_{DM}+\Omega_b+\Omega_r
	\ea
	where
		\ba
	\Omega_\phi=&-\sqrt{6} p_4 x+\frac{x^2 a_{2,Y}}{a_{4}}-\frac{y^2
		a_{2}}{2
		a_{4}}+(1-p_4) x^2\frac{a_{3}}{a_{4}}+
	\left(x^2 (p_2-p_4+2)+\sqrt{6}
	x\right)\frac{Y a_{3,Y}}{a_{4}}\\&+ \left(\sqrt{6} x (3
	p_2-5
	p_4+6)+4\right)\frac{Y a_{4,Y}}{a_{4}}+\left(2 \sqrt{6} x
	(p_2-p_4+2)+4\right)\frac{Y^2a_{4,YY}}{a_{4}}\\&+3(p_2-2
	p_4+1)\frac{Y
		a_{5} }{a_{4}}+\left(7 p_2-9 p_4+\frac{5
		\sqrt{\frac{2}{3}}}{3
		x}+12\right)\frac{Y^2a_{5,Y} }{a_{4}}\\&+ \left(2
	(p_2-p_4+2)+\frac{2 \sqrt{\frac{2}{3}}}{3
		x}\right)\frac{Y^3 a_{5,YY}}{a_{4}}\,.
	\ea
	The second Friedman equations reads
	\ba
	P_1\frac{dx}{dN}+P_2\frac{dy}{dN}+P=0
	\ea
	where
	\ba
	P=&1+\Omega_{r} w_r+x
	(p_2+p_4)\left(p_4 x+\sqrt{\frac{2}{3}}\right) +\frac{y^2 a_2}{2
		a_4}+(1-p_4) x^2\frac{
		a_3}{a_4}\\&-\left(x^2 (p_2-p_4+2)
	(p_2+p_4)+2 \sqrt{\frac{2}{3}} x
	(p_2+p_4)+2\right)\frac{Y
		a_{4,Y} }{a_4}\\&-(p_2-2
	p_4+1) \left(\sqrt{\frac{2}{3}} x (p_2+p_4)+1\right)\frac{Y
		a_5}{a_4}\\&
	-\left(\sqrt{\frac{2}{3}} x
	(p_2-p_4+2) (p_2+p_4)+\frac{2}{3} (2
	p_2-p_4+3)+\frac{2 \sqrt{\frac{2}{3}}}{3
		x}\right)\frac{
		Y^2 a_{5,Y}}{a_4}\,,
	\ea
	and
	\ba
	xP_1&=-\sqrt{\frac{2}{3}} x \frac{Y
		a_{3,Y}}{a_{4}}+ \left(-2 \sqrt{\frac{2}{3}} x
	(p_2-p_4+2)-\frac{8}{3}\right)\frac{Y^
		2 a_{4,YY}}{a_4}\\&
	+ \left(-\frac{4}{3}
	(p_2-p_4+2)-\frac{2 \sqrt{\frac{2}{3}}}{3
		x}\right)\frac{
		Y^3 a_{5,YY}}{a_{4}}+ \left(-\frac{4}{3} (3 p_2-4
	p_4+5)-\frac{\sqrt{\frac{2}{3}}}{x}\right)\frac{Y^2
		a_{5,Y}}{a_4}\\&+
	\left(\sqrt{\frac{2}{3}} x (-3 p_2+5
	p_4-6)-\frac{4}{3}\right)\frac{Y a_{4,Y}}{a_{4}}-
	(p_2-2 p_4+1)\frac{4 Y a_{5}}{3
		a_{4}}+\sqrt{\frac{2}{3}} p_4 x\,
	\ea
	and
	\ba
	yP_2&=\sqrt{\frac{2}{3}} x \frac{Y
		a_{3,Y}}{a_{4}}+\left(2 \sqrt{\frac{2}{3}} x
	(p_2-p_4+2)+\frac{8}{3}\right)\frac{Y^
		2 a_{4,YY} }{a_4}+
	\left(\sqrt{\frac{2}{3}} x (3 p_2-5
	p_4+6)+\frac{8}{3}\right)\frac{Y a_{4,Y}}{a_{4}}\\&
	+ \left(\frac{4}{3}
	(p_2-p_4+2)+\frac{2 \sqrt{\frac{2}{3}}}{3
		x}\right)\frac{
		Y^3 a_{5,YY}}{a_{4}}+ \left(\frac{14 p_2}{3}-6
	p_4+\frac{5 \sqrt{\frac{2}{3}}}{3
		x}+8\right)\frac{Y^2
		a_{5,Y}}{a_{4}}\\&+
	(p_2-2
	p_4+1)\frac{2 Y
		a_{5} }{a_{4}}-\sqrt{\frac{2}{3}}
	p_4 x-\frac{2}{3}\,.
	\ea

	The time derivative of the first Friedman equation reads
	\ba
	F_1\frac{dx}{dN}+F_2\frac{dy}{dN}+F=0\,,
	\ea
	where
	\ba
	F=\Omega_b \left(-\sqrt{6} p_2 x+\sqrt{6} q
	x-3\right)-\Omega_{DM} \left(\sqrt{6} p_2
	x+3\right)-\Omega_r \left(\sqrt{6} p_2 x+3
	w_r+3\right)\,,
	\ea
    and
	\ba
	xF_1&=2 x^2  \frac{Y
		a_{2,YY}}{a_{4}}+\frac{x^2
		a_{2,Y}}{a_{4}}+ \left(2 x^2
	(p_2-p_4+2)+2 \sqrt{6}
	x\right)\frac{Y^
		2 a_{3,YY}}{a_{4}}+(2-2 p_4) x^2
	\frac{a_3}{a_{4}}\\&+ \left(x^2 (4 p_2-6
	p_4+10)+3 \sqrt{6}
	x\right)\frac{Y
		a_{3,Y}}{a_{4}}
	+
	\left(4 \sqrt{6} x
	(p_2-p_4+2)+8\right)\frac{Y^3
		a_{4,YYY}}{a_{4}}\\&+ \left(4 \sqrt{6} x (4
	p_2-5
	p_4+8)+24\right)\frac
	{Y^2 a_{4,YY}}{a_{4}}
	+\left(\sqrt{6} x (9 p_2-17
	p_4+18)+6\right)\frac{Y
		a_{4,Y} }{a_{4}}\\&+ \left(4
	(p_2-p_4+2)+\frac{4 \sqrt{\frac{2}{3}}}{3
		x}\right)\frac{Y
		^4 a_{5,YYY}}{a_{4}}+(p_2-2
	p_4+1)\frac
	{6 Y a_{5} }{a_{4}}-\sqrt{6} p_4 x\\&
	+\left(26 p_2-30
	p_4+\frac{20 \sqrt{\frac{2}{3}}}{3
		x}+48\right)\frac{Y^3
		a_{5,YY} }{a_{4}}+\left(34 p_2-48 p_4+\frac{5
		\sqrt{\frac{2}{3}}}{x}+54\right)\frac{Y^2
		a_{5,Y} }{a_{4}}\,
	\ea
	and
	\ba
	yF_2&=-2 x^2 \frac{Y
		a_{2,YY}}{a_{4}}-\frac{x^2
		a_{2,Y}}{a_{4}}+ \left(-2 x^2
	(p_2-p_4+2)-2 \sqrt{6}
	x\right)\frac{Y^
		2 a_{3,YY}}{a_{4}}\\&+\left(x^2 (-4 p_2+6
	p_4-10)-4 \sqrt{6}
	x\right)\frac{Y
		a_{3,Y} }{a_{4}}
	+2 (p_4-1) x^2
	\frac{a_3}{a_{4}}\\&+\left(-4 \sqrt{6} x
	(p_2-p_4+2)-8\right)\frac{Y^3
		a_{4,YYY} }{a_{4}}+\left(-2 \sqrt{6} x (9
	p_2-11
	p_4+18)-32\right)\frac
	{Y^2 a_{4,YY} }{a_{4}}\\&
	+\left(-2 \sqrt{6} x (6 p_2-11
	p_4+12)-14\right)\frac{Y a_{4,Y} }{a_{4}}+ \left(-4
	(p_2-p_4+2)-\frac{4 \sqrt{\frac{2}{3}}}{3
		x}\right)\frac{Y^4 a_{5,YYY}}{a_{4}}\\&
	+ \left(-30 p_2+34
	p_4-\frac{26 \sqrt{\frac{2}{3}}}{3
		x}-56\right)\frac{Y^3
		a_{5,YY}}{a_{4}}+ \left(-48 p_2+66
	p_4-\frac{10
		\sqrt{\frac{2}{3}}}{x}-78\right)\frac{Y^2
		a_{5,Y}}{a_{4}}\\&-12 (p_2-2
	p_4+1)\frac
	{ Y a_{5}}{a_{4}}+2 \sqrt{6} p_4 x+2\,.
	\ea
	
	For the particular case $a_{4,Y}=a_{4,YY}=a_{4,YYY}=0$ we find for the background quantities
	\ba
	P_{1s}=\sqrt{\frac{2}{3}}p_4\quad,\quad P_{2s}=\frac{1}{3y_s}\left(3\frac{p_4}{p_2}-1\right)\,,
	\ea
	\ba
	F_{1s}=\frac{1}{\sqrt{6}}\left(p_4+3\frac{p_4^2}{p_2}-2p_2\Omega_\phi-\frac{9a_{2,YY}}{p_2^3y^2a_4}\right)\quad{\rm and }\quad F_{2s}=-\frac{1}{2y}\left(2\left(\Omega_\phi-2\right)-3\frac{p_4^2}{p^2_2}+5\frac{p_4}{p_2}+\frac{9a_{2,YY}}{p_2^4y^2a_4}\right)\,.
	\ea
	For the perturbations of the functions $P$ and $F$ we get
	\ba
	P_{x,s}&=\frac{1}{\sqrt{6}p_2}\left(3p_4\left(p_2-p_4\right)+2p_2^2\left(1-\Omega_{\phi}\right)\right)
	\quad{\rm ,}\quad
	x_sP_{x,s}+y_sP_{y,s}=-1+\frac{p_4}{p_2}
	\ea
	and
	\ba
	 	F_x=\sqrt{6}p_2\left(\Omega_{\phi}-1\right)\,.
	\ea
	\section{Perturbations}\label{app:perturbations}
Here we present for completeness the equations from Ref.~\cite{DeFelice:2011bh}. We have
\ba
\omega_1&\equiv 2\left(G_4-2XG_{4,X}\right)-2 X \left(\dot\phi H G_{5,X}-G_{5,\phi}\right)\\
\omega_2&\equiv -2G_{3,X}X\dot\phi+4G_4H-16X^2G_{4,XX}H+4\left(\dot\phi G_{4,\phi X}-4 H G_{4,X}\right)+2G_{4,\phi}\dot\phi-4 \dot\phi H^2 X^2 G_{5,XX}\\&-10 \dot\phi
H^2 X G_{5,X}+8 H X^2
G_{5,\phi X}+12 H X G_{5,\phi}\\
\omega_3&\equiv 3X\left(G_{2,X}+2XG_{2,XX}\right)+6X\left(3X\dot\phi H G_{3,XX}-G_{3,\phi X}X- G_{3,\phi}+6H\dot\phi G_{3,X}\right)\\&+18H\left(4HX^3G_{4,XXX}-HG_{4}-5X\dot\phi G_{4,\phi X}-G_{4,\phi}\dot\phi+7HXG_{4,X}+16HX^2G_{4,XX}-2X^2\dot\phi G_{4,\phi XXX}\right)\\&+6 H^2 X \left(2 \dot\phi H X^2 G_{5,XXX}+13 \dot\phi H X G_{5,XX}+15
\dot\phi H G_{5,X}-6 X^2
G_{5,\phi XX}-27 X G_{5,\phi X}-18 G_{5,\phi}\right)\\
\omega_4&\equiv 2 G_4-2 \ddot\phi X G_{5,X}-2 X
G_{5,\phi}
\ea
Then
\ba
c_T^2=\frac{\omega_4}{\omega_1}\quad,\quad Q_T=\frac{w_1}{4}
\ea
\ba
c_s^2=\frac{3\left(2\omega_1^2\omega_2H-\omega_2^2\omega_4+4\omega_1\omega_2\dot\omega_1-2\omega_1^2\dot\omega_2\right)-6\omega_1^2\rho_{DM}-6\omega_1^2\rho_{b}-6\omega_1^2(1+w_{\rm rad})\rho_{\rm rad}}{\omega_1\left(4\omega_1\omega_3+9\omega_2^2\right)}
\ea
and
\ba
Q_s=\frac{w_1\left(4w_1w_3+9w_2^2\right)}{3w_2^2}\,.
\ea
The no-ghost condition reads $c_T^2, c_s^2, Q_T, Q_s >0$. Let us first consider the tensor modes no-ghost conditions in general. We have that
\ba
Q_T=\frac{1}{2}\phi^{p_4}\left(a_4-2Y
	a_{4,Y}+p_5Ya_5-Y^2a_{5,Y} \left(
	(p_2-p_4+2)+\sqrt{\frac{2}{3}}\frac{1}{x}\right)\right)
\ea
and
\ba
c_T^2=\frac{
	a_4-p_5 Y
	a_5-Y^2a_{5,Y}  \frac{1}{\sqrt{6}x}\frac{d\ln Y}{dN}\,.
}{a_4-2Y
	a_{4,Y}+p_5Ya_5-Y^2a_{5,Y} \left(
	(p_2-p_4+2)+\sqrt{\frac{2}{3}}\frac{1}{x}\right)
}
\ea
Since the scalar sector is rather involved here we only present the formulas for our particular model where $a_3=a_5=a_{4,Y}=a_{4,YY}=0$. In this case we find
\ba
Q_s&=\frac{16}{3w_2^2y_s^2}\phi^{2p_4+p_2}{a_4^3(Y_s)}\Bigg(x_s\left(2p_4-p_2\right)\left(\sqrt{6}+3p_4x_s\right)-3\Omega_{DM}-3\Omega_{b}-3\Omega_{r}+\frac{x_s^2a_{2,YY}}{a_4}\\&
+\frac{12Y_s^3a_{4,YYY}}{a_4}\left(2+\sqrt{6}\left(2+p_2-p_4\right)x_s\right)\Bigg)
\ea
and
\ba
c^2_s&=\frac{16}{3w_2^2Q_sy_s^2}\phi^{2p_4+p_2}{a_4^3}\Bigg(x_s\left(2p_4-p_2\right)\left(\sqrt{6}+3p_4x_s\right)-3\Omega_{DM}-3\Omega_{b}-3\Omega_{r}\left(1+w_r\right)\Bigg)\,.
\ea

\bibliographystyle{jhep}
\bibliography{scaling_bib.bib}

\end{document}